\font\subtit=cmr12
\font\name=cmr8
\input harvmac

\def\plb#1#2#3#4{#1, {\it Phys. Lett.} {\bf {#2}}B (#3), #4}
\def\npb#1#2#3#4{#1, {\it Nucl. Phys.} {\bf B{#2}} (#3), #4}

\def\cmp#1#2#3#4{#1, {\it Comm. Math. Phys.} {\bf {#2}} (#3), #4}
\def\lmp#1#2#3#4{#1, {\it Lett. Math. Phys.} {\bf {#2}} (#3), #4}

\def\ijmpa#1#2#3#4{#1, {\it Int. Jour. Mod. Phys.} {\bf A{#2}} (#3), #4}
\def\jmp#1#2#3#4{#1, {\it Jour. Math. Phys.} {\bf {#2}} (#3), #4}

\def\jgp#1#2#3#4{#1, {\it Journal Geom. Phys.} {\bf {#2}} (#3), #4}

\def\fph#1#2#3#4{#1, {\it Fortschr. Physik} {\bf {#2}} (#3), #4}

\null\vskip 3truecm
\centerline{\subtit
QUANTUM FIELD THEORIES ON ALGEBRAIC CURVES\foot{Talk delivered by
F. Ferrari at the IX Max Born Symposium, Karpacz (PL), September 24--28,
1996.}
}
\vskip 1truecm
\centerline{F{\name RANCO} F{\name ERRARI}$^{a}$ and J{\name AN} T.
S{\name OBCZYK}$^b$}
\smallskip $^a${\it
LPTHE, Universit\'e Pierre and Maria Curie--Paris VI and
Universit\'e Denis Diderot--Paris VII, Boite 126, Tour 16, 1$^{er}$ \'etage,
4 place Jussieu, F-75252 Paris Cedex 05, France.
}
\smallskip $^b${\it
Institute for Theoretical Physics, Wroc\l aw
University, pl.  Maxa Borna 9, 50205 Wroc\l aw, Poland, E-mail:
jsobczyk@proton.ift.uni.wroc.pl}
\smallskip
\vskip 4cm
\centerline{ABSTRACT}
{\narrower \abstractfont
In this talk the main features of the operator formalism
for the $b-c$ systems on general algebraic curves developed in refs.
\ref\naszej{\jmp{F.  Ferrari, J.  Sobczyk and W.  Urbanik}
{36}{1995}{3216};
\ijmpa{F. Ferrari and J. Sobczyk}{11}{1996}{2213}.}--\ref\naszed{
\jgp{F. Ferrari and J. Sobczyk}{19}{1996}{287}.}
are reviewed. The first part of the talk is an introduction to the
language of algebraic curves. Some explicit
techniques
for the construction of meromorphic
tensors are explained. The second part is dedicated to the
discussion of the $b-c$ systems.
Some new results concerning the concrete representation
of the basic operator algebra of the $b-c$ systems and the
calculation of divisors on algebraic curves have also been
included.
} 
\pageno=0
\vfill\eject
\newsec {INTRODUCTION}
\vskip 1cm

In this talk some aspects of quantum field theories on Riemann surfaces
(RS) represented as algebraic curves are discussed.
The latter are defined as $n-$sheeted branched covers of the complex sphere
${\bf CP}^1$.
Any compact and orientable Riemann surface
can be represented in this way \ref\grha{
P. Griffiths
and J. Harris, {\it Principles of Algebraic
Geometry}, John Wiley \& Sons, New York 1978.}.
Until now, most of the physical literature deals with the
cases of hyperelliptic or $Z_n$ symmetric curves (see e.g. \ref\hyel{
\npb{D.  Lebedev and A.  Morozov}{302}{1986}{163}; \plb
{E. Gava, R. Iengo and G. Sotkov}{207}{1988}{283}; \plb{A. Yu. Morozov and
A. Perelomov}{197}{1987} {115}; \ijmpa{M.
A.  Bershadsky and A.  O.  Radul}{2}{1987}{165}.}),
while our curves are general.
One advantage of working on algebraic curves is the possibility of
constructing meromorphic tensors with poles at given points
in an explicit way. 
For this reason and also in view of possible future applications
in string theories and N=2 supersymmetric Yang--Mills
field theories \ref\witten{\npb{N. Seiberg and E. Witten}{426}{1994}{19}.},
the techniques for the derivation of
meromorphic tensors on general algebraic curves
are explained in Section 2 in great detail.
In Section 3 we explain the operator formalism developed in  refs.
\naszej--\naszed\
for the $b-c$ systems \ref\fms{
\npb{D. Friedan, E. Martinec and S.
Shenker}{271}{1986}{93}.}.
The basis used here for expanding the $b-c$ fields
on RS are similar to multipoint generalizations
\ref\genkn{\fph{R. Dick}{40}{1992}{519};
\lmp{M. Schlichenmaier}{19}{1990}{151}.}
of the Krichever--Novikov bases
\ref\kn{
I. M. Krichever and S. P. Novikov, {\it
Funk. Anal. Pril.} {\bf 21} No.2 (1987), 46;
{\bf 21} No.4 (1988), 47;
J. Geom. Phys. {\bf 5} (1988) 631.
}.
On the other
hand, our formalism has some analogies with that of ref.
\ref\blmr{\cmp{L. Bonora, A. Lugo, M. Matone and J. Russo}
{123}{1989}{329}.
}.
It is understood that the
parametrization of the moduli space of RS
is given by the branch points of the curve.
This choice entails some limitations, for instance in dealing
with fermions \foot{An exception is provided by the hyperelliptic
curves.}
or if the entire moduli space of the RS
of a given genus $g$ should be considered.
Nevertheless, such limitations are compensated by several advantages.
First of all, the explicitness with
which analytic field theories can be treated.
For instance, the meromorphic tensors involved in the correlation functions
of the $b-c$ fields can simply be written as rational functions of the
coordinates describing the algebraic curve.
Moreover, it is possible to treat
field theories on $n-$sheeted branched covers of
${\bf CP}^1$
as multivalued field theories on the complex sphere.
In our operator formalism, for example, all the correlation functions
of the $b-c$ systems are computed exploiting simple normal ordering
rules between operators defined on  ${\bf CP}^1$.
A concrete representation of the
Fock space is provided
in terms of semiinfinite forms at the end of Section 3.
Another important point is that the
multivalued field theories corresponding to the
$b-c$ systems defined on algebraic curves
can be reformulated as nonstandard conformal
field theories on the complex sphere. For some simple curves, these conformal
field theories have been explicitly constructed together with their
primary fields, see refs. \naszej--\naszed\
and references therein.
It has been shown that they
obey a complicated nonabelian statistics.
Furter possible developments of our results will be briefly
 discussed in the Conclusions.
\newsec {THE LANGUAGE OF ALGEBRAIC CURVES}

In order to fix the notations, let {\bf C} be the
complex plane and {\bf CP}$^1$ the complex projective line
which coincides
with the sphere or, equivalently, with the compactified complex
plane {\bf C}$\cup\{\infty\}$.
In this talk, we will consider RS as $n-$sheeted branched covers of
${\bf CP}^1$ \grha. The latter are algebraic curves defined as the locus
of points $(z,y)\in {\bf CP}^1\otimes {\bf CP}^1$ for which the
following equation is satisfied:
\eqn\curve{F(z,y)=0}
Here $F(z,y)$ denotes the so-called Weierstrass polynomial (WP),
which is of the form:
\eqn\weipd{F(z,y) = P_n(z)y^n + P_{n-1}(z)y^{n-1} +
\ldots + P_1(z)y + P_0(z) = 0}
with
$P_s(z)=\sum\limits_{m=0}^{n_s}\alpha_{s,m}z^m$ for $s=0,\ldots,n$
and $n_s\in{\bf N}$.
By a well known theorem, any Riemann surface can be expressed
as an algebraic curve of this kind.
Thus, from now on, we will use the words RS, 
$n-$sheeted branched covers of
${\bf CP}^1$ and algebraic curves interchangeably, though this language is
somewhat imprecise.
The best known algebraic curves are the
hyperrelliptic curves, whose WP is simply given by:
$y^2=-P_0(z)$.
Also the slightly more general $Z_n$ {symmetric
curves} will be often mentioned here.
They are characterized by a WP
of the kind $y^n=-P_0(z)$.

At this point, we solve eq. \curve\ with respect to $y$.
The resulting function $y(z)$ is multivalued on ${\bf CP}^1$
and exchanges its branches at a set of $N_{bp}$ branch points
$a_i,\ldots,a_{N_{np}}\in {\bf C}$ that will be defined below.
To simplify our analysis, we assume that $P_n(z)=1$ in eq. \weipd.
There is no loss of generality in this assumption.
As a matter of fact, if $P_n(z)\ne 1$, we perform in eq. \curve\ the change of
variables:
\eqn\firtra{\tilde y(z)=y(z)P_n(z)}
This does not affect the monodromy properties of $y(z)$, so that both
$\tilde y(z)$ and $y(z)$ are meromorphic functions on the same Riemann surface.
However,
it is easy to realize that $\tilde y$ satisfies
the equation
$\sum_{i=0}^n\tilde P_{n-i}(z) \tilde y^{n-i}=0$,
where now $\tilde P_n(z)=1$ and $\tilde
P_{n-i}(z)=P_{n-i}(z)P_n^{i-1}(z)$ for
$i=1,\ldots,n$.

Further, we suppose that none of the branch points is located at
in $z=\infty$.
The presence of a branch point at infinity can be detected trying the ansatz
\eqn\ansatz{y(z)_{\sim\atop z\rightarrow\infty}\gamma z^p+{\rm lower}\enskip
{\rm  order} \enskip {\rm terms}}
in eq. \curve\ for large values of $z$.
Solving eq. \curve\ at the leading
order, one determines the values of $\gamma$ and $p$.
A branch point at $z=\infty$ is indicated by noninteger values of $p$.
If this is the case, it is always possible without any loss
of generality to perform a birational
transformation on the curve of the following kind. 
First of all, the branch point at infinity is moved to a finite region
of the plane by exploiting an
$SL(2,{\bf C})$ transformation in $z$.
Of course, in doing this the condition $P_n(z')$ is spoiled in the
new variable $z'$, but can easily be restored with the
aid of the transformations \firtra\ in $y$.

We are now ready to define the branch points $a_1,\ldots,a_{N_{bp}}$.
Supposing that the curve \curve\ is nondegenerate\foot{See for instance
ref. \naszed\
for the definition of nondegeneracy.}, they are the solutions of
of the following system of equations:
\eqn\bpdef{F(z,y)=F_y(z,y)=0}
where $F_z(z,y)=dF(z,y)/dz$.
It is useful to eliminate from eq. \bpdef\
the variable $y$.
As an upshot, one obtains a polynomial equation in $z$
of the kind $r(z)=0$. Apart from very special curves,
in which $r(z)$ has multiple roots,
its degree coincides with the number of branch points $N_{bp}$.
Let us notice that it is possible to derive the resultant $r(z)$
of eqs. \bpdef\ explicitly using the dialitic
method of Sylvester, ref.
\ref\cenr{F.  Enriques and O.
Chisini, {\it Lezioni sulla Teoria
Geometrica delle Equazioni e delle Funzioni Algebriche}, Zanichelli,
Bologna (in italian).}, Vol. II, pag. 79.
To each branch point $a_i$ one can associate an integer $\nu_i$, called
the ramification index and defined as the number of branches of $y(z)$
that are exchanged at that branch point.
Clearly $\nu_i\le n$.
At a branch point of
ramification index $\nu_i$,  the WP $F(z,y)$
vanishes together with its first $n-1$ partial derivatives
in $y$. 
The genus $g$ of the Riemann surface \curve, the ramification indices
of the branch points and the number of sheets $n$ composing the curve
are related together by the Riemann--Hurwitz formula:
\eqn\rihu{2g-2 = -2n + \sum_{s=1}^L (\nu_s -1)}
The genus can be explicitly computed once the form of the Weierstrass
polynomial is known exploiting the Baker's method, see
ref.
\ref\forsyth{A. R. Forsyth, {\it Theory of Functions of a Complex
Variable}, Vols I and II, Dover Publications, Inc., New York, 1965.},
Vol. I,
pag. 404 and will not be reported here.

On a Riemann surface $S$ represented as an $n-$sheeted cover of ${\bf
CP}^1$ there is a ``canonical'' complex structure inherited from
${\bf CP}^1$. A possible atlas on $S$ is the following.
Let us put $R={\rm max}|a_i|$ and $\rho={\rm min}|a_i-a_j|$ for
$i,j=1,\ldots,N_{bp}$. Near a branch point $a_i$ of ramification index
$\nu_i$, or more precisely in the open disk $|z-a_i|<\rho$, we choose
the local coordinate $\xi^{\nu_i}=z-a_i$.
For $|z|>R$, the local coordinate is $z'=1/z$. Let us notice
that on the algebraic curve the set $|z|>R$ corresponds to an union of
$n$ disjoint discs.
On the remaining open sets that build the covering of $S$ the 
local coordinate is $z$.

To conclude this Section, we discuss the meromorphic tensors and their
divisors. In particular, we are interested in tensors of the kind
$T^{(l)}(z)dz^\lambda$, with $\lambda$ upper or lower indices
depending on the sign of $\lambda=0,\pm 1,\pm 2,\ldots$
The meromorphic functions correspond to the case $\lambda=0$.
The branch index $l$ is to recall that a tensor $T$ is
in general multivalued on ${\bf CP}^1$ due to its
dependence in $y(z)$.
Let $[D]$ a
finite collections of points $p_1,\ldots,p_r$ with multiplicities
$l_1,\ldots,l_r$ (integers) on the RS. $[D]$ is a so-called divisor
\ref\fakra{H. Farkas and I. Kra, {\it Riemann Surfaces},
Springer Verlag, 1980}.
To any meromorphic tensor $Tdz^\lambda$ with zeros $z_r$ of order $k_r$
and poles $p_s$ of order $l_s$, one can associate a divisor $[T]$
of the kind:
\eqn\div{[T] = \sum_r k_rz_r - \sum_s l_sp_s}
The most general tensor on an algebraic curve can be written in the
form:
\eqn\protodiff{T^{(l)}(z)dz^\lambda= Q(z,y^{(l)}(z)){dz^\lambda\over
[F_y(z,y^{(l)}(z))]^\lambda}}
where $Q(z,y)$ is a rational function of $z$ and $y$.
The reason for which the factor $[F_y(z,y^{(l)}(z))]^{-\lambda}$
has been singled out in \protodiff\ will be clear below.
From eq. \protodiff\ it is evident that, in order to construct tensors
on an algebraic curve with poles and zeros at given points, it is necessary
to know at least the divisors of the basic building blocks $dz$, $y$ and
$F(z,y)$. This can be done quite explicitly for general WP \weipd\ if
$P_n(z)=1$ and there are no branch points at infinity.
We only need the additional assumption that the polynomials
$P_1(z)$ and $P_0(z)$ appearing in the WP have no roots in common.
In this way, 
eq. \curve\ is approximated  for small values of  $y$
by the
relation
$y\sim -P_0(z)/P_1(z)$.
Therefore, the zeros $q_1,\ldots,q_{n_0}$ of $y(z)$ occur for values of
$z$ corresponding to the roots of
$P_0(z)$.
To study the behavior of $y(z)$ at infinity we try the ansatz
\ansatz\ in eq. \curve. If we retain only the leading order terms
of $y(z)$ and of the polynomials $P_s(z)$ appearing in the WP \weipd,
then eq. \curve\ is approximated by:
\eqn\leaord{\gamma^n
z^{pn}+\alpha_{s,n_s}\ldots+\gamma^{n-s}z^{p(n-s)+n_s}+
\ldots+\alpha_{0,n_0}z^{n_0}=0 }
Since $y(z)$ is not branched
at infinity, there should be
$n$ different solutions for  $\gamma$ that satisfy \leaord.
Clearly, this can be true only if 
the first and last monomials $\gamma^nz^{pn}$ and $\alpha_{0,n_0}z^{n_0}$
entering in eq. \leaord\
are of the same
order near $z=\infty$, i. e. $z^{pn}\sim z^{n_0}$.
Moreover, all the other monomials must not contain higher order powers in $z$.
Thus, we obtain for $p$ the following result:
\eqn\aku{p={n_0\over n}=1,2,\ldots}
so that $n_0$ is an integer multiple of $n$.
In this way we have derived the divisor of $y$:

\eqn\divy{ [y]=\sum^{n_0}_{r=1}q_r\ -\ \sum^{n-1}_{j=0}{n_0\over n}\infty_j }

where the symbols $\infty_j$ denote the points on the curve
corresponding to $z=\infty$.
As we see from the above equation, the degree of the divisor $[y]$ is
zero as expected for a meromorphic function.
Analogously, it is
possible to compute also the divisors of $F_y(z,y)$ and $dz$:
\eqn\divdf{[F_y]=\sum^{n_{bp}}_{r=1}(\nu_{r}-1)a_r\ -\
(n-1)\sum^{n-1}_{j=0}{n_0\over n}\infty_j }
\eqn\divdz{ [dz]=\sum^{n_{bp}}_{r=1}(\nu_r-1) a_r\ -\
2\sum^{n-1}_{j=0}\infty_j }
The details are explained in ref. \naszed.
Exploiting the above divisors, one is able to prove that,
if $\lambda\ge 2$, the following tensor has only a single pole at the
point $z=w$ on the sheet $l=l'$:
\eqn\weiker
{
K_{\lambda}^{(ll')} (z,w)dz^\lambda = {1\over z-w} 
{F(w,y^l(z))\over y^l(z) - y^{l'}(w)}
{dz^\lambda
 \over [F_y(z,y^l(z))]^{\lambda} }}
where the indices $l$ nd $l'$  label the branches in $z$ and $w$
respectively.
The tensor $K_{\lambda}^{(ll')} (z,w)dz^\lambda $ will be hereafter called
the Weierstrass kernel (WK).
If $\lambda =1$, it is easy to check that
\eqn\ake{\omega_{ww'}^{ll'l''}(z)dz = K_1^{(ll')} (z,w)dz-
K_1^{(ll'')} (z,w')dz}
is a differential of the third kind with two poles in $z=w$ and $z=w'$
on the sheets $l=l'$ and $l=l''$ respectively. 
Let us remember that the differentials of the first kind are the
holomorphic one
forms, the differentials of the second kind are meromorphic one
forms with no residua and the differentials of the third kind are 
meromorphic one forms with only single poles.
Finally, a nondegenerate metric on the curve is
$g_{z\bar z}dzd\bar z=(1+z\bar z)^\beta{dzd\bar z\over |F_y(z,y)|^2}$
$\beta = p(n-1) -2$.

To fix the ideas,
in the next Section we will consider only the 
class of curves in which the degree $n_s$ of the polynomials $P_s(z)$
appearing in eq. \curve\ is $n_s=n-s$. Let us denote these algebraic
curves with the symbol $\Sigma_g$. Their genus $g$ is given by the
formula:
\eqn\genus{g={(n-1)(n-2)\over 2}}
Moreover $N_{bp}= 2g+2(n-1)$ denotes the total number of simple
$\nu_i=2$ branch points.
Finally, the divisors  of $dz$, $F_y$ and $y$ correspond to the subcase
$n_0/n=1$ of
eqs. \divy--\divdf.
 
\newsec{THE OPERATOR FORMALISM FOR THE $b-c$ SYSTEMS}

On the curves $\Sigma_g$ defined above let us consider the theory of the
$b-c$ systems with spin $\lambda$. If $\xi,\bar \xi\in \Sigma_g$ is a set
of complex coordinates on the Riemann surface,
the variable $z\in {\bf CP}^1$ can be regarded
as a mapping $z:\xi\rightarrow{\bf CP}^1$. Thus,
putting $\bar \partial \equiv\partial/\partial \bar z$,
$b\equiv b^{(l)}(z(\xi), \bar z(\bar \xi))dz^\lambda $ and $c\equiv
c^{(l)}(z(\xi),  \bar z(\bar \xi))dz^{1-\lambda} $, it is possible
to write the action of the $b-c$
systems as follows:
\eqn\action{S_{\rm
bc}=\int_{\Sigma_g} d^2z(\xi)\left(b\bar\partial c+\bar b \partial\bar c
\right)}
>From now on, we will suppose that $\lambda\ge2$.
The case $\lambda=1$ is complicated due to the presence of extra zero
modes\foot{By zero modes we intend the global solutions of the
classical equations of motion.}
in the $c$ fields and will not be discussed here.
A detailed treatment of the operator formalism with $\lambda=1$
can be found in the second of refs. \naszej.

To expand the classical $b-c$ fields, we use the generalized Laurent series
(GLS) of ref. \naszed.
These consist in two different expansions for the fields $b$ and $c$ and
necessarily
contain modes which are multivalued on ${\bf CP}^1$:
\eqn\fkn{ 
f_{k,i}(z) =
{z^{-i-\lambda}y^{n-1-k}(z)dz^{\lambda}\over (F_y(z,y(z)))^{\lambda} }
}
$$
\phi_{l_i}(w) dw^{1-\lambda}
= { 
w^{-i+\lambda-1}
dw^{1-\lambda}\over(F_y(w,y(w)))^{1-\lambda}}\times
$$
\eqn\phikn{
\left( y^l(w)+y^{l-1}(w)P_{n-1}(w)+y^{l-2}(w)P_{n-2}(w)+...+P_{n-l}(w)
\right)
}
In ref. \naszed,
we have proved that any tensor \protodiff\ is a linear combination
of the modes \fkn-\phikn. Moreover, after exchanging $\lambda$ with
$1-\lambda$, the two bases \fkn\ and \phikn\ turn out to be equivalent,
i.e. the modes of the first can be expressed in terms of the modes of the
second.
The GLS for the $b-c$ fields read:
\eqn\bdzcdz{b(z)dz^\lambda=\sum_{k=0}^{n-1}b_k(z)dz^\lambda\qquad\qquad\qquad
c(z)dz^{1-\lambda}=\sum_{k=0}^{n-1}c_k(z)dz^{1-\lambda}}
where
\eqn\bkdz{b_k(z)dz^\lambda=
\sum\limits_{i=-\infty}^{\infty}b_{k,i}
f_{k,i}(z)dz^\lambda}
\eqn\ckdz{c_k(z)dz^{1-\lambda}=
\sum\limits_{i=-\infty}^\infty
c_{k,i}\phi_{k,i}(z)dz^{1-\lambda}}
The fields $b_k(z)dz^\lambda$
and $c_k(z)dz^{1-\lambda}$ have a physical significance.
After quantizing the theory, in fact, we will see that fields with different
values of $k$ do not interact.
This splitting of the $b-c$ systems into $n$ different ``$k-$sectors''
is already evident at the classical level.
For instance, it is possible to find a number $N_{b_k}$
of $b$ zero modes $\Omega_{k,i}(z)dz^\lambda$ that are proportional
to $f_k(z)dz^\lambda$. The zero modes are of the form
\eqn\zm{
\Omega_{k,i}dz^\lambda
=f_{k,i}(z)
dz^\lambda
}
The range of $i$ in the above equation
and the value of $N_{b_k}$ strongly depend on the WP.
The complete list of cases has been worked out in ref. \naszed\
and will not
be reported here.
For instance, if $n>4$ and 
$\lambda>1$ we have:
\eqn\caseone{\cases{k=0,\ldots,n-1\cr
\lambda(2-n)+n-1-k\le i\le -\lambda\cr
N_{b_k}=\lambda(n-3)+k-n+2\cr}}
It is possible to verify that
$\sum\limits_{k=0}^{n-1} N_{b_k}=(2\lambda-1)(g-1)$ 
as desired.
An important motivation for choosing the GLS
\bdzcdz-\ckdz\ is that the WK \weiker\ has a very simple expansion
in the modes \fkn\ and \phikn:
\eqn\keyfor{K_\lambda(z,w)=
{1\over z-w}\sum_{k=0}^{n-1}\phi_{k,1-\lambda}(w)f_{k,\lambda}(z)}
At this point we quantize the $b-c$ systems on the algebraic curve,
treating the theory as a set of $n$ noninteracting field theories
on ${\bf CP}^1$.
To this purpose, we treat
the
coefficients $b_{k,i}$
and $c_{k,i}$ appearing in the GLS as quantum operators, for which we postulate
the following commutation relations (CR):
\eqn\commrel{\{b_{k,j},c_{k',j'}\}=\delta_{kk'}\delta_{j+j',0}.}
The consistency of the above assumptions, suggested by the multivaluedness
of the modes \fkn-\phikn\ entering in the GLS \bdzcdz,
will be proved ``a posteriori''
by showing
that in this way the correct correlations functions of the $b-c$ systems are
obtained.
The operators carrying the index $k$ for
$k=0,\ldots,n-1$, act on the vacua $|0>_k$, 
where $|0>_k$ is the standard $SL(2,{\bf C})$ invariant vacuum of the complex
sphere.
The ``total vacuum" of the $b-c$ systems is
\eqn\totalvacuum{|0\rangle=\otimes_{k=0}^{n-1}|0\rangle_k}
Generalizing the usual definitions of creations and destruction
operators at genus zero, we demand that:
\eqn\ban{b^-_{k,i}|0\rangle_k\equiv b_{k,i}|0\rangle_k=0
\qquad\qquad\qquad\left\{\eqalign{k=&0,\ldots,n-1\cr i
\ge& 1-\lambda\cr}\right.}
\eqn\can{c^-_{k,i}|0\rangle_k\equiv c_{k,i}|0\rangle_k=0
\qquad\qquad\qquad\left\{\eqalign{k=&0,\ldots,n-1\cr i\ge&
\lambda\cr}\right.}
Moreover. we introduce the "out" vacua ${}_k\langle 0|$ requiring that
\eqn\bcrea{{}_k\langle0| b^+_{k,i}\equiv {}_k\langle 0|b_{k,i}=0
\qquad\qquad\qquad\left\{\eqalign{k=&0,\ldots,n-1\cr i
\le& -\lambda-N_{b_k}\cr}\right.}
\eqn\ccrea{{}_k\langle 0|c^+_{k,i}\equiv {}_k\langle 0|c_{k,i}=0
\qquad\qquad\qquad\left\{\eqalign{k=&0,\ldots,n-1\cr i\le&
\lambda-1\cr}\right.}
>From the above equations, we see
that some of the $b_{k,j}$'s correspond to zero modes and the remaining
ones are organized in two sets of creation and annihilation operators. 
The same applies to the $c_{k,j}$ 
with the only difference that there are no zero modes for them.
>From the above equations and \keyfor\ we deduce the following
natural definition of the ``normal ordering'' between the fields:
\eqn\bzcwnorm{b_k(z)c_k(w)dz^\lambda dw^{1-\lambda}
=:b_k(z)c_k(w):dz^\lambda dw^{1-\lambda}+
K_\lambda(z,w)dz^\lambda dw^{1-\lambda}}
The ``time ordering" is implemented by
the requirement that the fields $b(z)$ and $c(w)$
are radially ordered with respect to the variables $z$ and $w$.
Let us notice that, remarkably, there is no need of introducing
more complicate time
ordering which is sensitive to the branches of the $b-c$ fields. 
Finally,  in order to take into account the zero modes, we impose the following
conditions:
\eqn\vaccond{{}_k\langle 0|0\rangle_k=0\qquad{\rm if}\ N_{b_k}\ne 0;
\qquad\qquad
{}_k\langle 0|\prod\limits_{i=1}^{N_{b_k}}b_{k,i}|0\rangle_k=1.}
At this point, we have all the ingredients
to compute the correlation functions of
the $b-c$ systems within our operatorial formalism.
Here we report only the result for the propagator
in the case of $n>4$, $\lambda>1$:
$${\langle 0| b^{(l)}(z) c^{(l')}(w)\prod\limits_{I=1}^{N_b}
 b^{(l_I)}(z_I)
|0\rangle\over
\langle 0|\prod\limits_{I=1}^{N_b} b^{(l_I)}(z_I)|0\rangle}=$$
\eqn\propfin{{
{\rm det}\left|\matrix{\Omega_{1,1}^{(l)}(z)&\ldots&
\Omega_{n-1,N_{b_{n-1}}}^{(l)}(z)&K_\lambda^{(ll')}(z,w)\cr
\Omega_{1,1}^{(l_1)}(z_1)&\ldots&
\Omega_{n-1,N_{b_{n-1}}}^{(l_1)}(z_1)&K_\lambda^{(l_1l')}(z_1,w)\cr
\vdots&\ddots&\vdots&\vdots\cr
\Omega_{1,1}^{(l_{N_b})}(z_{N_b})&\ldots&
\Omega_{n-1,N_{b_{n-1}}}^{(l_{N_b})}(z_{N_b})&K_\lambda^{(l_{N_b}l')}(z_{N_b},w)
\cr}\right |\over
{\rm det}\left|\Omega_I(z_J)\right|}}
In the above equation,
$N_b$ denotes the total number of $b$ zero modes.
Let us notice that the right hand side of eq. \propfin\ is a ratio of
correlators containing
multivalued fields on thecomplex plane, whereas the left hand side represents
the propagator of the $b-c$ systems on the algebraic curve.
The detailed computations of the general $n-$ point functions of
the $b-c$ systems can be found in refs. \naszej--\naszed.

To conclude this Section,
some remarks should be made about the structure of the linear space on which
the $b-c$ systems are realized as quantum field theories.
A concrete
realization of the representation space can be given in terms of ref.
semiinfinite forms via the identifications
\eqn\realb{ b_{k,j} \hookrightarrow \beta_{k, j}
\wedge ...\qquad\qquad\qquad
c_{k,j} \hookrightarrow
{
\partial\over \partial
\beta_{k, -j}
}
}
The "in" vacuum state is filled with the excitations $\beta_{k, p_k}$ with
$p_k\geq 1-\lambda$.
For
\eqn\xinfstate{|x\rangle = \beta_{k,j_1}\wedge \beta_{k,j_2}\wedge ...}
and
\eqn\yinfstate{|y\rangle = \beta_{k,j_1'}\wedge \beta_{k,j_2'}\wedge ...}
one introduces the bilinear form
\eqn\biform{ \langle y| x\rangle \equiv ...\wedge \beta_{k,-j_2'+H}\wedge
\beta_{k,-j_1'+H}\wedge \beta_{k,j_1}\wedge \beta_{k,j_2}\wedge ... }
where
\eqn\przes{H = 1 - 2\lambda - N_{b_k}}
The above bilinear form
is by definition equal zero unless all the excitations are present
at the RHS of \biform . If not zero, it can take only the
values $\pm 1$ (some sign convention
has to be made). The vacuum state and the bilinear forms
introduced above satisfy the requirements \vaccond.
In the linear space of states
\eqn\lss{ b_{k_1,s_1} ... c_{l_1,t_1} ...|0\rangle }
it is natural to introduce another bilinear form
\eqn\scprod{ (y|x) \equiv ...\wedge \beta_{k,-j_2'+H}\wedge
\beta_{k,-j_1'+H}\wedge \beta_{k,j_1}\wedge \beta_{k,j_2}\wedge ... }
by demanding that \scprod\ is zero unless all the excitations except
from those corresponding to zero modes are present on the RHS. This time
$(0|0)=1$. We do not obtain a Hilbert space of states as $(\cdot |\cdot )$
is not
positive definite. This should not be a surprise as for instance
the $b-c$ systems for 
$\lambda=2$ are the Faddeev-Popov ghosts for reparameterization
invariance in string theory. Let us mention also that the consistency
of the above representation requires the following reality conditions
for the elementary excitations:
\eqn\con{b^+_{k,j} = b_{k,-j+H},\qquad c^+_{k,j} = c_{k,-j+H}
}
It is important that the set of $b_{k,j}$ corresponding to zero modes
of the theory remains invariant under the conjugation introduced in \con .
\newsec{CONCLUDING REMARKS}

As we have seen, any meromorphic tensor on general algebraic curves
can be expanded as linear combinations of the multivalued modes
\fkn--\phikn.
The operator formalism, instead, has been tested until now only in the
case of the $b-c$ systems.
However, one should try to generalize it also to other field theories.
For instance, the $\beta-\gamma$ systems seems to be treatable as well.
Also the theory of massless scalar systems is a good candidate for applying
our methods, but the fact that the correlation functions
are no longer meromorphic complicates
the study of this case.
Finally, we hope that our formalism could be useful in discussing the
minimal models on algebraic curves
\ref\arm{\plb{C. Crnkovic, G.M. Sotkov and M. Stanishkov}{220}{1989}{397};
S. A. Apikyan and C. J. Efthimiou, {\it Minimal Models of CFT
on $Z_N$ Surfaces}, hep-th 9610051.}.

\listrefs
\bye